\documentclass[12pt,preprint]{aastex}

\begin{document}
 
\title{The Unusual Luminosity Function of the Globular Cluster M10}

\author{Denise L. Pollard, Eric L. Sandquist, Jonathan R. Hargis}
\affil{San Diego State University, Department of Astronomy, San Diego,
CA 92182} \email{pollard@sciences.sdsu.edu, erics@sciences.sdsu.edu,
jhargis@sciences.sdsu.edu}

\author{Michael Bolte} \affil{University of California Observatories,
University of California, Santa Cruz, CA 95064}
\email{bolte@ucolick.org}

\begin{abstract}
We present the $I$-band luminosity function of the differentially
reddened globular cluster M10. We combine photometric analysis
derived from wide-field ($23\arcmin \times
23\arcmin$) images that include the outer regions of the
cluster and high-resolution
images of the cluster core. After making corrections for incompleteness
and field star contamination, we find that the relative numbers of
stars on the lower giant branch and near the main-sequence turnoff are
in good agreement with theoretical predictions. However, we detect
significant ($> 6 \sigma$) excesses of red giant branch stars above
and below the red giant branch bump using a new statistic (a
population ratio) for testing relative evolutionary timescales of 
main-sequence and red giant stars. The statistic is insensitive to assumed
cluster chemical composition, age, and main-sequence mass
function. The excess number of 
red giants cannot be explained by reasonable systematic
errors in our assumed cluster chemical composition, age, or 
main-sequence mass function. Moreover, M10 shows excesses when
compared to the cluster M12, which has nearly identical metallicity,
age, and color-magnitude diagram morphology. We discuss possible
reasons for this anomaly, finding that the most likely cause is a mass
function slope that shows significant variations as a function of
mass.

\end{abstract}

\keywords{stars: evolution --- stars: luminosity function --- 
globular clusters: individual (M10, M12)}

\section{Introduction}

The luminosity function (LF) of a star cluster contains a great deal of
information about the 
physics of stellar evolution, star formation, and galactic and
many-body stellar dynamics. The extraction of this information can be
complicated by this superposition of effects, as well as by purely
observational problems such as photometric crowding and incompleteness
in stellar counts. Counts
of post-main-sequence stars in star clusters contain information on
evolutionary timescales \citep{renz}, and through them, the physical
processes occurring in and near regions of nuclear fusion. Star counts
in the globular cluster M30 previously revealed a discrepancy
between theoretical predictions and observations of the relative
numbers of red giant branch (RGB) and main sequence (MS) stars
\citep{bolte,berg,guhath,sand99}. Similar studies of other globular
clusters (M5, \citet{sand96}; M3, \citet{rood}) have not found such
discrepancies. 

In this paper, we present the unusual LF of the
globular cluster M10 (NGC 6254; C1654--040). The M10 LF has previously
been derived from observations of much smaller fields by \citet{hurl}
and \citet{Piotto99}. Neither study had large enough samples
of cluster member
evolved stars to make useful comparisons with stellar evolution
theoretical models. \citet{Piotto99} did, however, find some
interesting features in their analysis of deep $HST$ observations near
the cluster half-mass radius: the M10 main-sequence LF was significantly
steeper than those of the clusters M22 and M55, and there was a
peculiar bump in the LF of the upper MS at $6.5 \lesssim M_V \lesssim
7.5$. In this paper we present the LF for a large number of evolved
stars in M10, and discuss some of the peculiar features of the evolved-star
LF in the context of the main-sequence star LF.

\section{Observations and Data Reduction}

\subsection{Wide-Field Data}

The primary observations for this study were made on the nights of 6
May 1995 and 9 May 1995 (UT) using the Kitt Peak National Observatory
(KPNO) 0.9 m telescope. In total, 10 images were obtained in $BVI$
filters (3 each in $B$ and $V$, and 4 in $I$). One 10s image and one
60s image were taken in each band on night 3 (6 May 1995) of the run,
with an additional 200s image in the $I$-band. On the photometric
night 6 (9 May 1995) of the run, an additional image was taken in each
band (a 10s exposure in $B$-band and a 60s exposure in both $V$- and
$I$-bands). Seeing conditions were similar to those cited in Hargis et
al. (2004) for the cluster M12. All data were taken using a $2048
\times 2048$ pixel CCD chip with a plate scale of 0\farcs68
pixel$^{-1}$, so that the total sky coverage was 23\farcm2 $\times$
23\farcm2 centered on the cluster.

\begin{figure}
\plotone{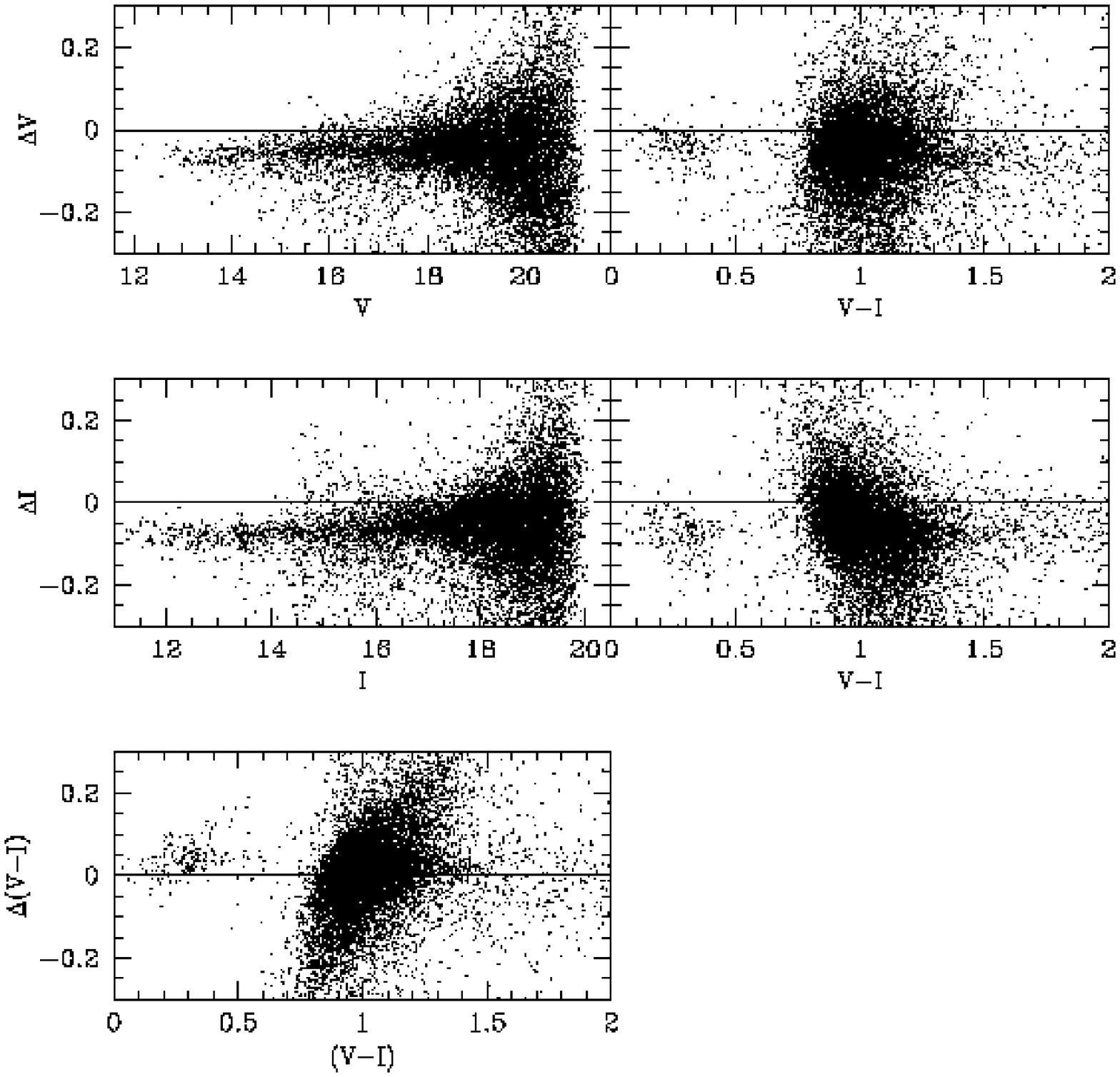}
\caption{Residuals (in the sense of ours minus theirs) from 13,000 stars
in common with \citet{Kaspar}.\label{resid}}
\end{figure}

The frames were processed in standard fashion using
IRAF\footnote{IRAF(Image Reduction and Analysis Facility) is
distributed by the National Optical Astronomy Observatories, which are
operated by the Association of Universities for Research in Astronomy,
Inc., under contract with the National Science Foundation.} tasks and
packages. The bias level was removed by subtracting a fit to the
overscan region and a master bias frame. Both twilight and dome flat
fields were used in constructing a master flat field frame from the
high spatial frequency component of the dome flats and the
low-frequency (smoothed) component of the twilight flats. The M10
profile-fitting photometry was performed using the DAOPHOT II/ALLSTAR
package of programs (Stetson 1987). The reduction and calibration procedures 
were very similar to those described in Hargis et
al. (2004) for M12, and used the same photometric calibration
fields. Therefore we will not repeat the details
here. Fig. \ref{resid} shows a comparison between our M10 photometry
and that of \citet{Kaspar}. We note that there are significant offsets
between the two datasets, in the sense that $V_{us} - V_{vB} = -0.08$
and $(V-I)_{us} - (V-I)_{vB} = 0.02$. We found similar systematic
differences in comparing our M12 dataset \citep{Hargis} with that of
von Braun, but no such offsets were found in comparisons between ours
and other datasets from the literature. Thus, we believe our
photometry is properly calibrated to the standard system.

We obtained $BVI$ photometry for nearly 25,000 stars reaching from the
tip of the red giant branch (RGB) to over 1.5 magnitudes below the
turnoff in $I$ from the KPNO images. Fig. \ref{cmds} shows our $VI$
photometry for a subsample eliminating the center of the cluster (to
eliminate many blended stars) and the outskirts (where field stars
start to dominate) in order to illustrate the photometric quality. The
photometry was corrected for differential extinction using the
extinction map from von Braun et. al. (2002), which noticeably reduced
the scatter around the cluster sequences in the color-magnitude diagram 
(CMD). We focus on the
$I$-band LF in order to minimize the residual effects of the
differential reddening. 

We determined the cluster fiducial line from the dereddened dataset
using a method similar to that of \citet{Hargis}. Main sequence points
were determined from the mode of stars in magnitude bins, while
subgiant branch stars were determined from the mode of stars in color
bins. On the red giant branch, fiducial points were determined from
means. The fiducial is provided in Table \ref{fidtab}.
\subsection{High-Resolution Data}

We reduced additional observations of the core of the cluster using
the High-Resolution Camera (McClure et al., 1989)
on the 3.6m Canada-France-Hawaii Telescope
(CFHT) on 13 April 1993. In all, there
were eleven $B$-band exposures ($1\times 60$s, $10\times 300$s) and 
fourteen $I$-band exposures ($2\times 6$s,
$12\times 160$s). The observations used a 1200 $\times$ 1200 pixel Loral 3 CCD
with 0\farcs11 pixel$^{-1}$ for a total field of 2\farcm2 $\times$
2\farcm2. The images were processed similarly to the KPNO images, with
the exception that only twilight flat images were employed.
The High-Resolution Camera used a closed-loop, fast tip-tilt correction 
system to obtain very good image quality. Seeing ranged from about 0\farcs5 (FWHM) to 0\farcs9 for the M10 images.

\begin{figure}
\plotone{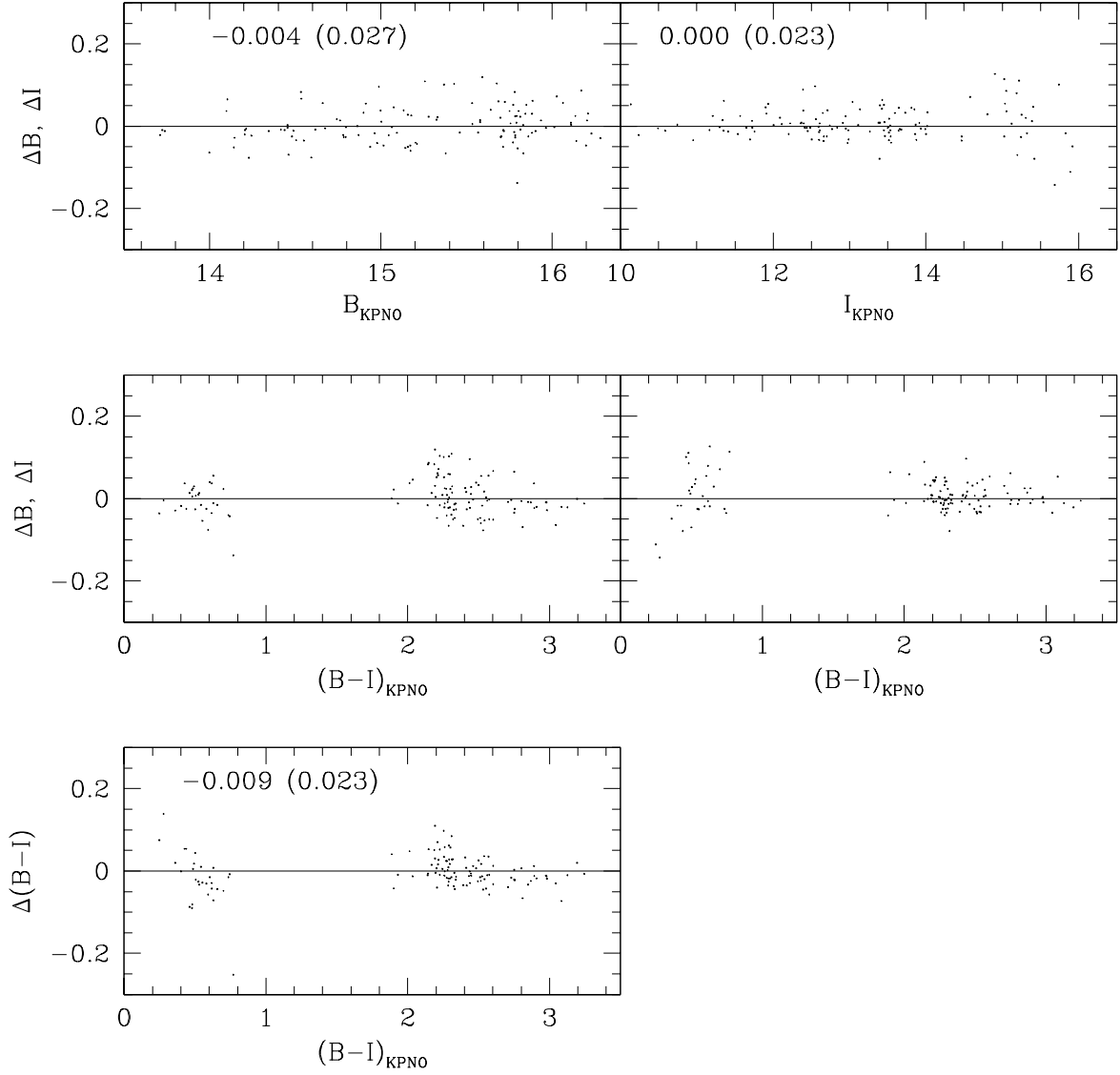}
\caption{Residuals (in the sense of CFHT minus KPNO) from 125 stars
used to calibrate the CFHT data.\label{cfhtcal}}
\end{figure}

\begin{figure}
\plotone{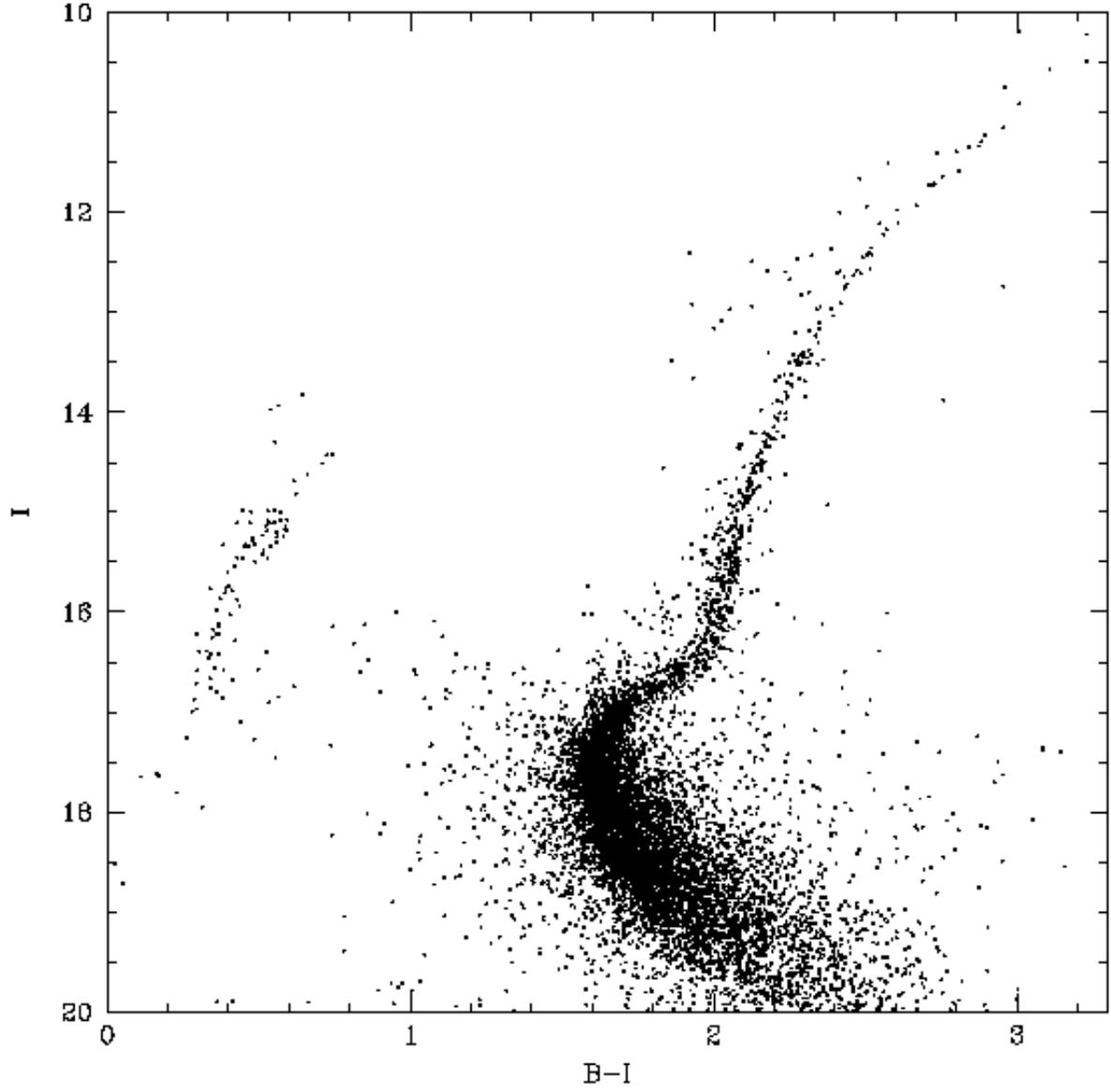}
\caption{($I,B-I$) color-magnitude diagram for the entire CFHT dataset.
\label{cfhtcmd}}
\end{figure}

The CFHT photometry was calibrated against KPNO photometry for 125
horizontal branch and bright giant branch stars. The transformation
equations used were
\[ b = B + a_{0,i} + a_1 * (B-I)\]
\[ i = I + b_{0,i} + b_1 * (B-I)\]
where $a_{0,i}$ and $b_{0,i}$ are zero-point corrections determined
for each CFHT image. Fig. \ref{cfhtcal} shows the residuals for the
calibration. The average residuals were less than 0.01 mag in $B$,
$I$, and $(B-I)$. No significant color trends were noticeable among
the red giant stars, although there may be a small trend among the
horizontal branch stars. Because the horizontal branch stars do not
impact our luminosity function considerations, we have not pursued the
issue. The final color-magnitude diagram is shown in Fig. \ref{cfhtcmd}.

\subsection{Artificial Star Tests and Completeness Corrections}

\begin{figure}
\plotone{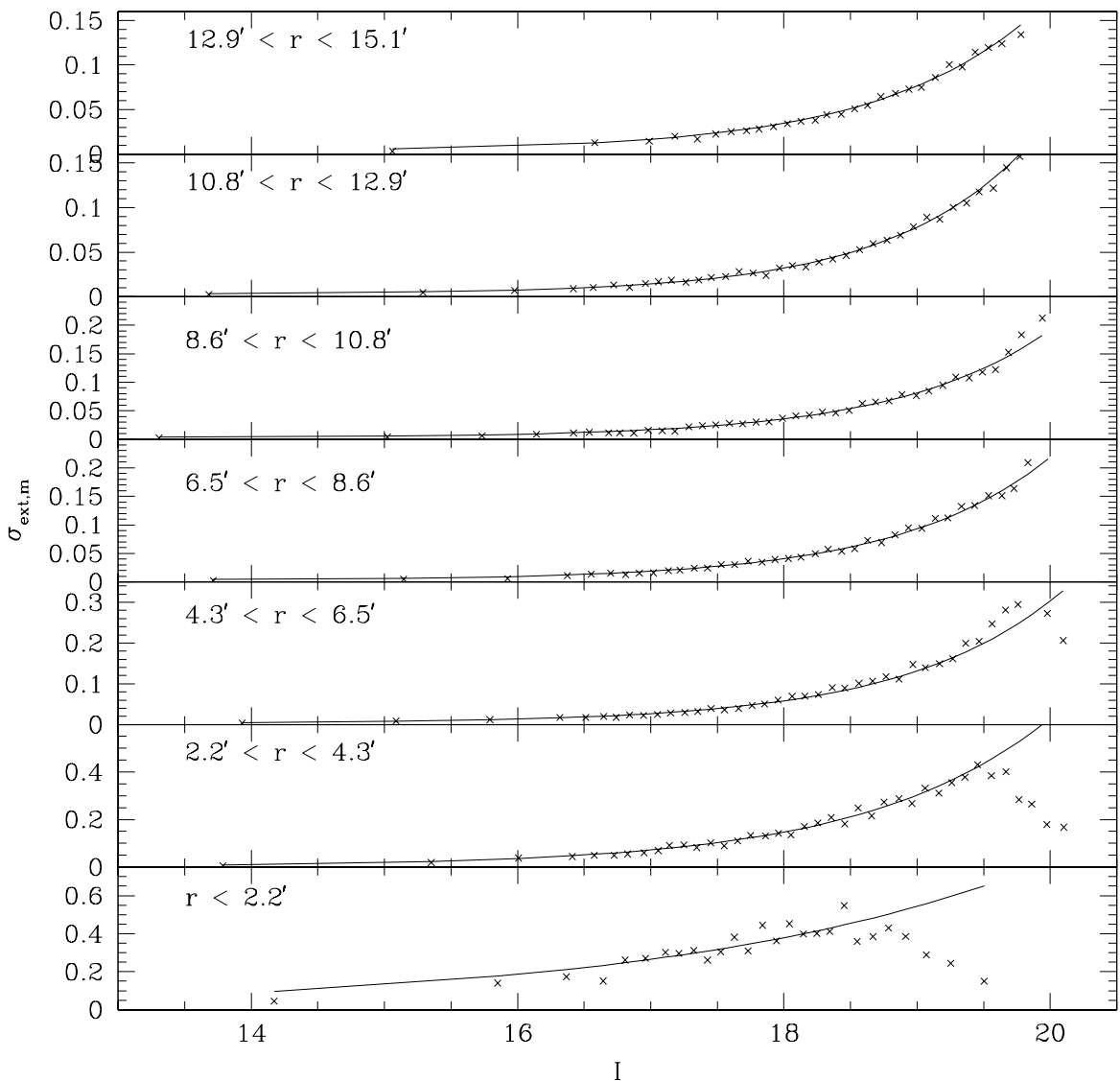}
\caption{Results from the artificial star tests for the external $I$
magnitude errors $\sigma_{ext}(I)$ as a function of radius and input magnitude.
\label{sige}}
\end{figure}

\begin{figure}
\plotone{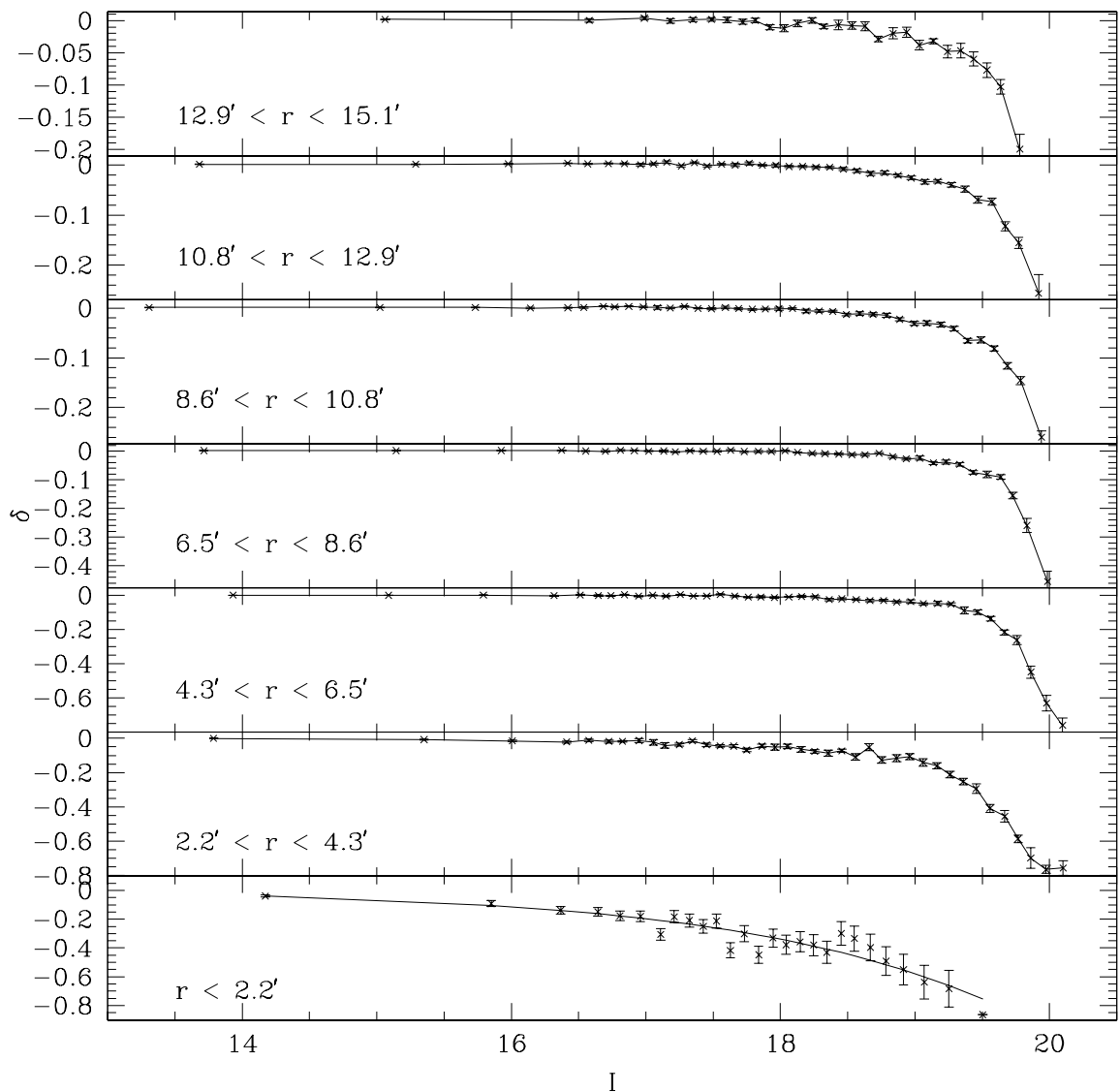}
\caption{Results from the artificial star tests for the $I$
magnitude biases $\delta(I)$ as a function of radius and input magnitude.
\label{delt}}
\end{figure}

\begin{figure}
\plotone{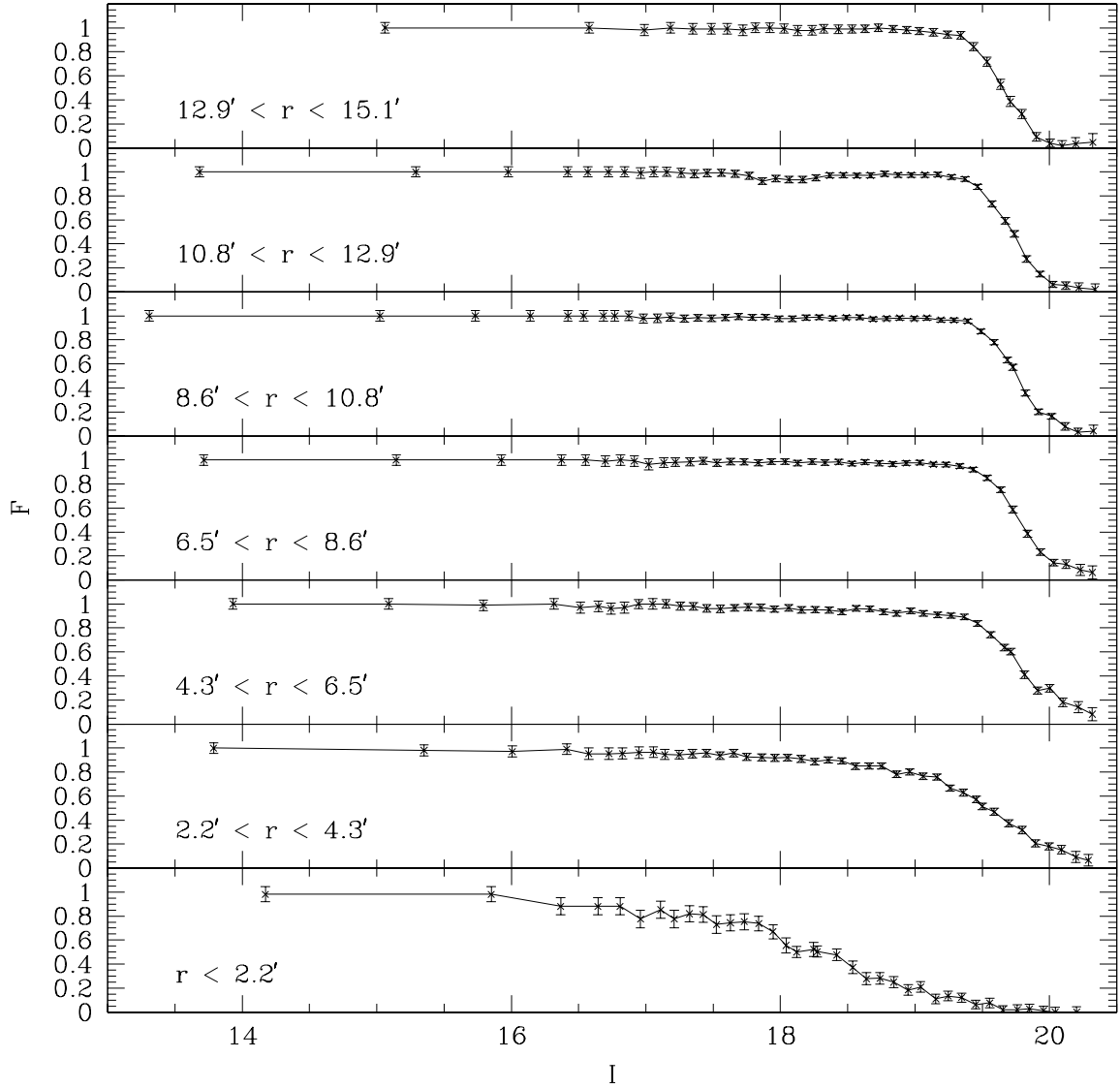}
\caption{Results from the artificial star tests for the total recovery
probability $F$ as a function of radius and input magnitude.
\label{bigF}}
\end{figure}

To quantify the completeness in detecting and counting stars
as a function of magnitude and position in
the cluster, we performed extensive artificial star tests on the KPNO
images following the same procedure described by Hargis et
al. (2004). One modification made for this study was the addition of
appropriate differential reddening for each artificial star according
to its position in the frame.  Positions for the artificial stars were
chosen at random within a grid such that no artificial stars
overlapped [separations were no smaller than 2 $\times$ PSF radius + 1
pixel; \citet{Piotto99}]. Artificial star tests were only conducted on
the $V$ and $I$ frames, with approximately 2000 stars added per frame
per run. The artificial star frames were processed in a manner
identical to our initial photometric reduction.  In all, we conducted
56 runs involving 91757 artificial stars.  The artificial star tests
were used to compute median color and magnitude biases $\left[
\delta_I \equiv \mbox{median}(I_{output} - I_{input}) \right]$, median
external color and magnitude error estimates $\left[ \sigma_{ext}(I)
\equiv \mbox{median}\mid\delta_I - \mbox{median}(\delta_I)\mid /
0.6745 \right]$, and total recovery probabilities $\left[ F(I)
\right]$.  These quantities are used to correct the LF for
incompleteness and magnitude biases \citep{sand96}, so we must be able
to find their values as a function of magnitude and position. We
fitted the external magnitude errors using functional forms given in
\citet{sand96}.  We found that linear interpolation for $\delta$ and
$F$ as a function of magnitude produced an improved description of the
behavior at the faint end of the sample. A relatively small number of
artificial stars were placed in the innermost radial bin due to its
small area, so we resorted to fits using functions from
\citep{sand96}.  Radial interpolation was accomplished using
polynomial functions.  We present the results for $\sigma_{ext}$,
$\delta$, and $F$ as a function of input artificial star magnitude and
radius in Figs. \ref{sige} - \ref{bigF}.

\begin{figure}
\plotone{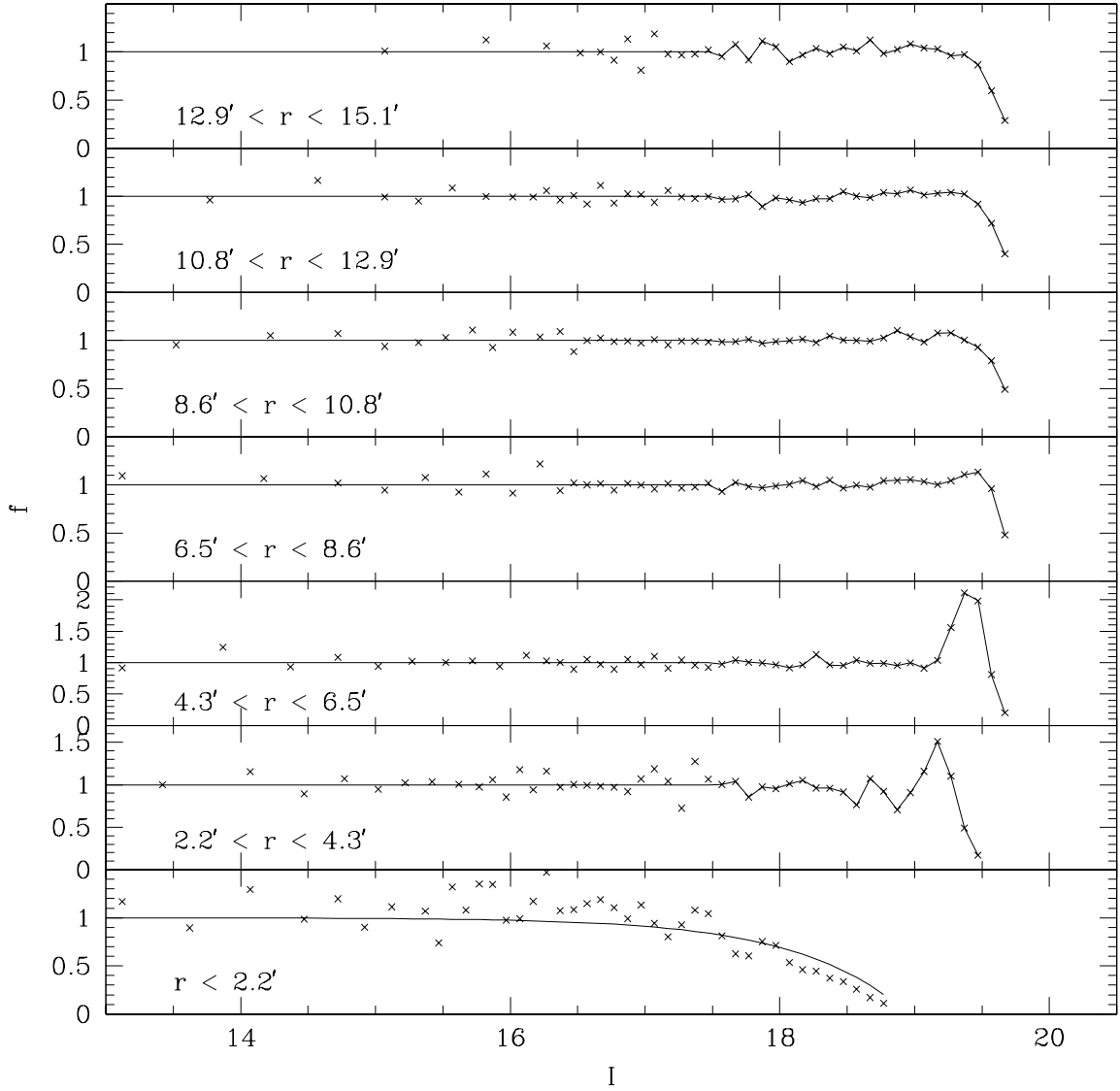}
\caption{Results from the artificial star tests for the completeness
fraction $f$ as a function of radius and output magnitude.
\label{fcomp}}
\end{figure}

The completeness corrections $f$ were subsequently computed following
the procedure of \citet{sand96}. $f$ was set to 1 for stars brighter
than the cluster turnoff to minimize numerical noise in the final
LF. Fig. \ref{fcomp} show the results for the completeness corrections
as a function of output magnitude and radius. In two of the radial
bins close to the cluster core, blending causes $f$ to become
substantially greater than 1. This fact drove our decision to limit
the luminosity function to $I < 19$ even though the photometry is
nearly 100\% complete for fainter stars farther from the cluster
center.

\subsection{Field Star Correction}

A significant field star population covers the cluster field, and the
field stars overlap the cluster fiducial in the CMD on the main sequence,
subgiant branch, and lower giant branch. The primary source of
contamination is from foreground main sequence stars in the Galactic
disk. We attempted to minimize field star contamination by
eliminating stars more than 9\farcm6 from the cluster center. We used
stars more than 11\farcm3 from the center to determine field star
corrections for the LF. As can be seen in the CMD of Fig. \ref{cmds},
the cluster star population is not readily apparent in the outer parts
of the image. To determine the field correction, we determined the
contribution to the LF from the outer portions of the field, and
multiplied it by a factor of 1.97 to account for the difference in
areal coverage. This correction will be an overestimate of the true
field star correction because we have not accounted for the small
population of cluster stars falling in the outer parts of the images.

\subsection{The Luminosity Function}

\begin{figure}
\plotone{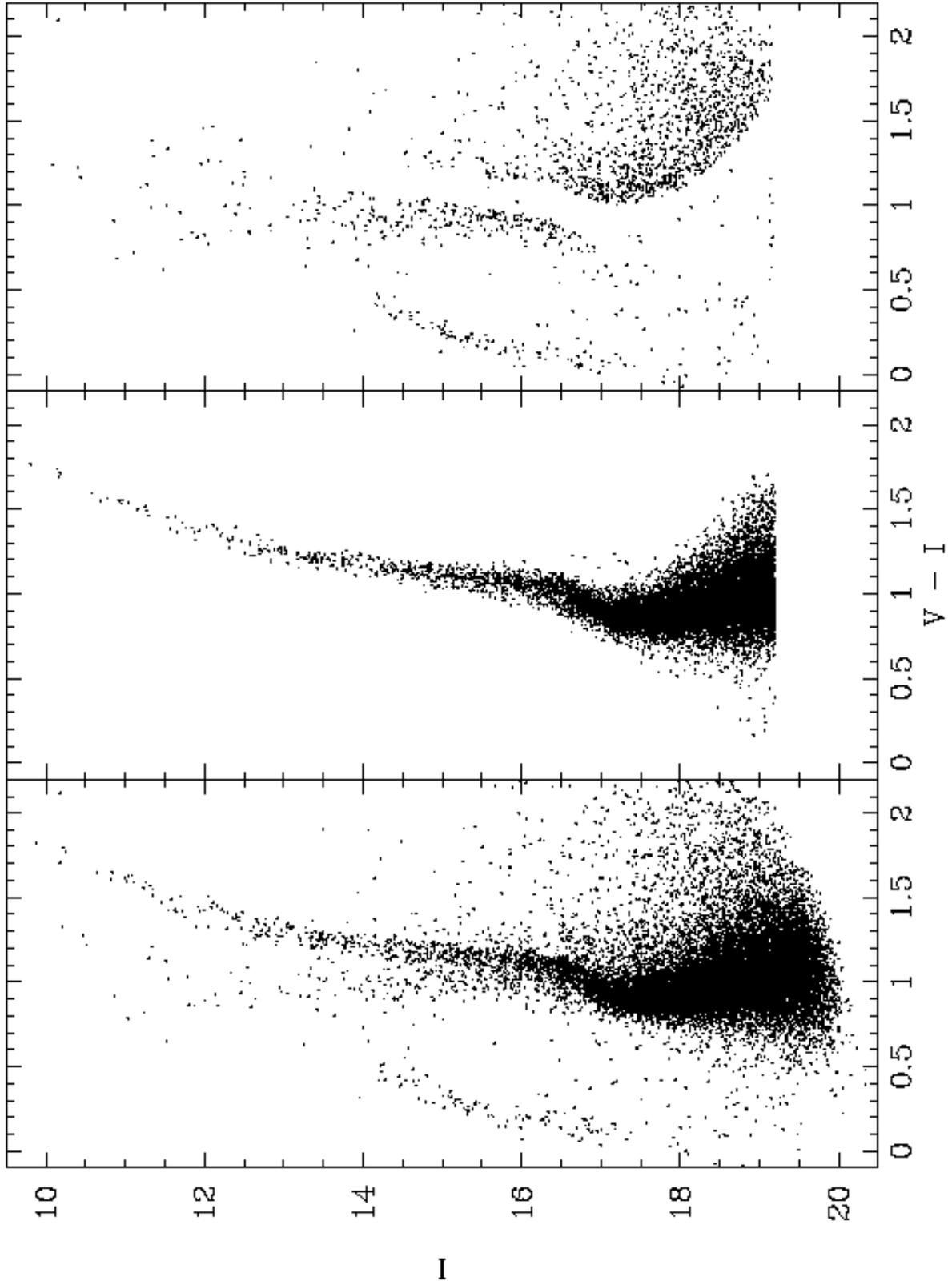}
\caption{{\it Left panel:} Dereddened CMD for measured M10 stars with
$r < 2\farcm3$ from the cluster center. {\it Middle and right panels:}
Stars that were and were not selected for the LF above the faint limit
($I = 19.2$).
\label{keptlost}}
\end{figure}

Using the results of the artificial star tests (particularly the
completeness factor $f$), we computed the observed LF following the
procedure of \citet{sand96}. The LF for the KPNO data was calculated by
multiplying each star detected on the object images by the factor
$f^{-1}$, which was tabulated as a function of magnitude and radius.
We selected stars for the LF that were less than $5 \sigma_{ext}$
(color and magnitude distance) from the fiducial line determined for
the cluster (Fig. \ref{keptlost}).  Because of the effects of crowding
near the cluster center in the KPNO images, we did not consider
main-sequence stars within 170\arcsec of the cluster center or evolved
stars within 68\arcsec. A correction was applied to the evolved star
sample to account for the differences in spatial coverage
\citep{sand96}.

The field-star-corrected portion of the LF is shown in
Fig. \ref{fieldcorr}. The corrected and uncorrected $I$-band
luminosity functions are presented in Table \ref{lftab}. Notable
corrections (relative to the size of the sample from the central
regions of the cluster) were in the bins at $I = 16.37$ and 16.57.
The field correction here significantly changes the shape of the LF at
the join between the SGB and the lower RGB. Smaller corrections are
needed in brighter bins on the RGB (with the exception of one bin at
$I = 13.97$, which becomes consistent with 0 to within the errors when
corrected). No field stars are found in the regions of the CMD
populated by RGB stars in the red giant bump ($I \sim 13.5$) and
brighter because the giant branch slopes away from the disk main
sequence population. We will return to the subject of the field star
contamination in \S \ref{lfsec}.

\subsubsection{The Combined Luminosity Function}

We produced a ``global'' LF combining data from the
KPNO and CFHT images in order to test the reality of features on the
RGB. We have broken the LF into three magnitude ranges:

BRIGHT ($I < 15.07$): For bright RGB stars, we took photometry from
the CFHT images for the center of the cluster and from the KPNO images
for stars with $r > 68\arcsec$. Extinction differences between the CFHT
field and the reference field of \citet{Kaspar} were also accounted
for. There is a small amount of area near the core that is not covered
by either dataset. A small number of stars were measured in both, but
were only counted once for the final LF. AGB stars can be clearly
distinguished from RGB stars in both samples, so there is minimal
contamination.

MIDRANGE ($15.07 < I < 17.67$): KPNO stars were only used if $r >
170\arcsec$, and incompleteness corrections from the artificial star
tests were also employed. CFHT stars were used for $r > 16\farcs5$,
but no incompleteness corrections were applied. 

FAINT ($I > 17.67$): Only KPNO stars with $r > 170$ were used and
incompleteness corrections were applied.

The BRIGHT and MIDRANGE segments of the combined LF were normalized to
samples with $r > 170\arcsec$ using stars in the ranges $13.07 < I
< 15.07$ and $15.07 < I < 16.07$, respectively.

In the lower panel of Fig. \ref{fieldcorr}, we compare the
field-star-corrected KPNO, CFHT, and combined LFs for the cluster. The
three agree very well, indicating that there is little or no radial
variation in the RGB populations. We will return to this issue in \S
\ref{disc}.

\section{M10 Characteristics: Age, Metallicity, Reddening, and 
Distance Modulus}

Before we compare the observed LF with theoretical models, we discuss
the accepted ranges for cluster characteristics (age, metallicity,
reddening zeropoint, and distance modulus).

The latest studies of the cosmic background radiation data from WMAP
have found the age of the universe to be 13.7$\pm$0.2 Gyr
\citep{Spergel2003}, setting a tight upper limit on the possible ages
of GGCs. Using the \textit{relative} age indicator $\Delta V
^{HB}_{TO}$ (defined as the magnitude difference between the $V$
magnitude of the ZAHB and MSTO points), \citet{Rose99} (which uses the
homogeneous data set presented in \citet{Rose00}) found a value of
$\Delta V^{HB}_{TO}$ = 3.50$\pm$0.11 for M10. The $\Delta V
^{HB}_{TO}$ value indicates that M10 is coeval with the oldest
globular clusters (to within the measurement errors), so that we will
only consider absolute ages between 11 and 13 Gyr.  Theoretical models
that include helium diffusion \citep{yy} adequately model the cluster
CMD with this constraint, but we have also compared the observed LF
with models \citep{bv} that do not include helium diffusion and which
require us to assume that the cluster is older than this range of ages.

There have been a number of determinations of [Fe/H] for M10, and
published values range over about 0.4 dex. The two most widely used
metallicity scales are those of \citet{ZW84} and \citet{CG97}, and we
considered both in our comparisons. Zinn \& West (ZW) cite a value of [Fe/H]
= $-1.60$. Using high-resolution spectra of GGC red giants, Carretta
\& Gratton (CG) measured [Fe/H] $= -1.41 \pm 0.02$. Systematic differences
between the two scales are well-documented, with the CG scale giving a
higher metallicity by approximately $0.2-0.3$ dex for low- or
intermediate-metallicity clusters. Spectroscopic measurements of the
infrared Ca II triplet of M10 red giants have also been made by
\citet{R97}. \citet{RHS97} used these measurements to compute
abundances on the ZW and CG scales of [Fe/H]$_{ZW} = -1.55 \pm 0.04$
and [Fe/H]$_{CG} = -1.25 \pm 0.03$. Recent high-resolution spectra
taken by \citet{KI03} were also used to measure
a metallicity [Fe/H]$_{KI}$ between $-1.41$ and $-1.48$ (dependent
on the model atmospheres used in the analysis).

In this study, we adopt the reddening values determined by
\citet{Kaspar}. We note, however, that their mean reddening value of
E$(V - I)$ = 0.28 disagrees with the value E$(V - I)$ = 0.39 derived
from COBE infrared dust emissivity maps by \citet{SFD98}. Although we
do not directly use the mean E$(V-I)$ value, the reddening issue is a
significant source of systematic uncertainty in doing absolute
comparisons with theoretical LFs, and so it is further discussed
below.

We computed the distance modulus via subdwarf fitting to the main
sequence photometry of \citet{Kaspar} because our photometry did not
reach faint enough on the MS to be useful for this purpose. We used
$Hipparcos$ parallaxes and subdwarf metallicities from \citet{carsbd}
as discussed in \citet{Hargis}. Relative to von Braun's differentially
dereddened data, $(m-M)_V = 14.18 \pm 0.04$ for our ``best"
assumptions: the reddening zero point E$(V-I)$ = 0.23 \citep{Kaspar},
and [Fe/H]=$-1.41$ from \citet{CG97}. (Note that the reddening zero
point is {\it not} the same as the average reddening. The reddening
zero point is the average reddening of a $280\arcsec \times
280\arcsec$ subfield chosen as a reference by von Braun et al. for
their larger M10 field.) By far the uncertainty in the reddening zero
point is going to be the largest systematic error. A 0.01 mag error in
E$(V-I)$ gives roughly a 0.054 mag error in the distance
modulus. Unfortunately the systematic error in the zero point cannot
be determined well.  Reddening maps from COBE data \citep{SFD98}
indicate E$(V-I)$ = 0.34 for the reddening zero point. The COBE zero
point can effectively be ruled out given that i) it implies a distance
modulus of 14.72, ii) M10 and M12 appear to be coeval with nearly
identical metallicities, and iii) the difference between the distance
moduli of M10 and M12 is no more than about 0.2 mag based on the
CMD. As evidence of this we present the fiducial lines of the two
clusters in Fig.\ref{fidcomp}. The data from the two clusters were
taken during the same observing run, and calibrated nearly
identically.

However, based on the COBE results, it is more likely that the
reddening zero point is larger than that of \citet{Kaspar} rather than
smaller. Thus we consider a distance modulus range $(m-M)_V =
14.18^{+0.15}_{-0.06}$ set by our estimation of the systematic
reddening uncertainties. When incorporating the systematic difference
between von Braun's data and ours (his is fainter than ours by 0.08
mag on average), this gives $(m-M)_V = 14.10^{+0.15}_{-0.06}$, which
is consistent with the distance estimates from \citet{Ferraro99} using
the horizontal branch. Using the Ferraro et al. value of E$(B-V) =
0.28$ and $A_V/{E(B-V)} = 3.1$, we calculate values of $(m-M)_V^{CG97}
= 14.25$ and $(m-M)_V^{[M/H]} = 14.22$ from their tables. By way of
comparison, the distance modulus for M12 derived via an almost
identical procedure was $(m-M)_V^{M12} = 14.05\pm0.12$ (assuming
[Fe/H]$_{M12} = -1.41$ and E$(V-I)_{M12} = 0.25 \pm 0.02$).


\section{The Luminosity Function}\label{lfsec}

\subsection{Comparison with Models}

\begin{figure}
\plotone{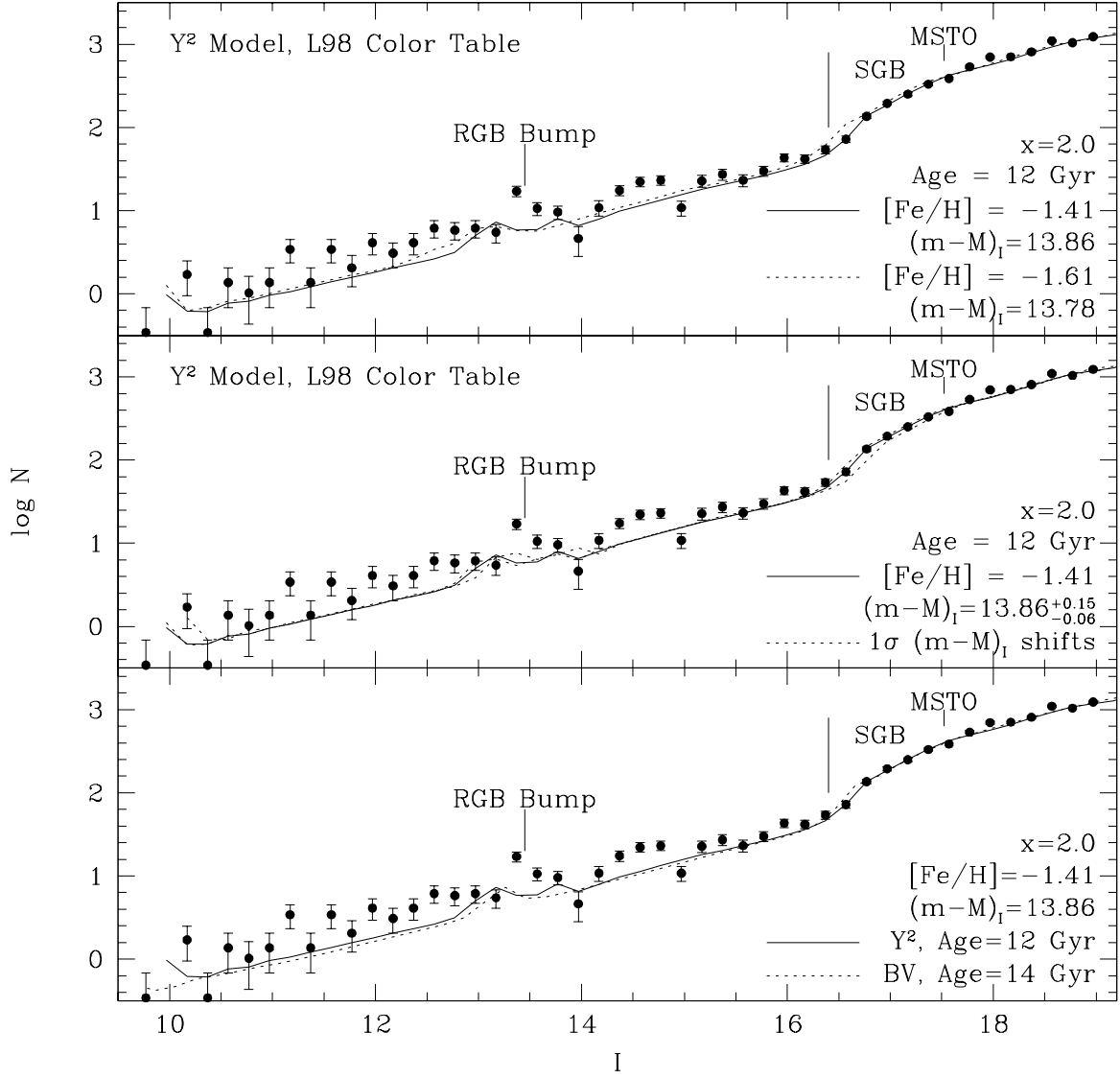}
\caption{Comparisons between the observed LF (combined sample) of M10
and theoretical models.  {\it Top panel:} models with varying
[Fe/H]. {\it Middle panel:} models with varying distance modulus. {\it
Bottom panel:} models with ``best'' choices for [Fe/H], distance
modulus, and ages from \citet{bv} and \citet{yy}.\label{lfcomp}}
\end{figure}

In Fig.~\ref{lfcomp}, we compare the observed luminosity function of
M10 with theoretical models covering realistic ranges in [Fe/H] and
distance modulus, and with different sets of models from \citet{yy}
and \citet{bv}. The models are normalized on the main sequence below
the turnoff (at $I \sim 18.5$), with the mass function exponent $x = 2$
chosen to match the slope of the main sequence LF. Models with varying
age are not plotted because within the acceptable age range there is
little difference in the degree of agreement between models and
observations. For realistic choices of age, [Fe/H], and distance
modulus, we find that the number of stars on the lower RGB relative to
main sequence stars is in good agreement with models. The RGB bump is
detected at $I \sim 13.3$.

There are a couple of deviations from the theoretical models that
deserve additional discussion.
There is a significant secondary bump at $14 < I < 15$. There is an
increase in counts in this range for both the KPNO and CFHT datasets
individually, and field star corrections do not remove the feature.
There also seems to be a significant excess in counts brighter than
the RGB bump in the range $12.2 < I < 13.2$. In Fig. \ref{lfcomp},
this excess is masked somewhat by the theoretical predictions for the
RGB bump. (We will not discuss the RGB bump in detail, however.) On a
final note, the main sequence mass function exponent ($x \approx 2$)
is relatively large as well, in agreement with the results of
\citet{Piotto99}.

To gauge the significance of the apparent excesses of RGB stars
relative to main sequence stars in Fig.~\ref{lfcomp}, we devised a
ratio of the number of stars in a magnitude range on the RGB to the
number near the cluster turnoff.  This selection is based on the
finding of \citet{stetlf} that cluster luminosity functions very
nearly overlie one another (independent of [Fe/H], [$\alpha$/Fe], age,
and initial mass function; see also \citet{larson}) when they are
shifted so that the turnoffs are coincident in magnitude.  By defining
samples of cluster stars relative to easily-measured points on the
cluster's fiducial line and taking the ratio of samples, we can nearly
eliminate uncertainties resulting from imperfectly known cluster
parameters and from the shifting and normalization of the theoretical
luminosity functions. In addition, statistical errors are easy to
determine using Poisson statistics.

We define the magnitude range of the RGB sample ($N_{RGB}$) relative
to the turnoff magnitude $I_{TO}$ ($17.52 \pm 0.10$ for M10). The
turnoff sample $N_{TO}$ is selected from the stars in a 0.4-magnitude
wide bin centered on $I_{MSTO}$. Because our final luminosity function
employs corrections for field contamination and incompleteness, we
computed the ratio from the LF. For the count excess in $14 < I < 15$,
we find a ratio $N_{RGB} / N_{TO} = 0.116 \pm 0.006$, and for the
excess with $12.2 < I < 13.2$, $N_{RGB} / N_{TO} = 0.0354 \pm
0.002$. The major source of systematic uncertainty is the measurement
of $I_{TO}$; if it shifts with respect to the RGB magnitude range
being used, the ratio value changes substantially. If $I_{TO} =
17.42$, the two ratios become 0.130 and 0.0397, and they become 0.103
and 0.0312 if $I_{TO} = 17.62$.

Using theoretical models, we can compute expected values for the
ratios. From the Yonsei-Yale isochrones \citep{yy}, we find ratios of
0.0694 and 0.0227 for [Fe/H]$ = -1.41$, [$\alpha$/Fe] = 0.3, age 12
Gyr, and mass function exponent $x=2.0$. Because our reference
population is at the turnoff and not fainter, the ratios are
insensitive to large changes in the mass function exponent. Reducing
the mass function exponent to $x = -0.5$ only increases the ratios to
0.0771 and 0.0254, respectively --- stellar evolution, rather than
star formation processes, dominates the ratio.  Between the ages
of 10.5 and 13.5 Gyr, the theoretical predictions for
$N_{RGB}/N_{MSTO}$ change by only 0.0095 and 0.0056, respectively.
Over the metallicity range $-1.61
\leq \mbox{[Fe/H]} \leq -1.41$, the theoretical predictions for
the two ratios change by only 0.0089 and 0.0013, respectively.
Thus, no reasonable variation of parameters like age or [Fe/H] or
possible errors in the determination of $I_{MSTO}$ is able to account
for the excess of RGB stars in the observed LF of M10. Comparing our
best observational value and the best theoretical model, the
differences are significant at more than the $7\sigma$ and $6\sigma$
levels for the samples fainter than ($14 < I < 15$) and brighter than
($12.2 < I < 13.2$) the RGB bump.

\subsection{Comparison with M12}

In addition to comparing the LF to theoretical models, it is also
valuable to compare to other globular clusters.  The cluster M12
provides an excellent comparison. Using relative age indicators, we
find that the ages of M10 and M12 are consistent to within measurement
errors. The color differences between the turnoff and red giant branch
(see Fig. \ref{fidcomp}) and the magnitude differences between turnoff
and horizontal branch ($\Delta V^{HB}_{TO} = 3.60 \pm 0.12$ for M12)
are both close for the two clusters. The metallicities are also close:
the ZW values only differ by 0.01 dex, while \citet{RHS97}
measurements indicate a difference of 0.15 dex or less with M12 the
more metal-rich (consistent with its redder upper RGB).
The distances of the two clusters are also the same to within the
measurement errors. One notable difference between the clusters is the
main sequence mass function exponent $x$: \citet{Hargis} found $x = 0$
for M12, while we find $x = 2$ for M10 in this study.

The differences in mass function mean that the two LFs cannot be
compared simply by correcting for differences in distance modulus and
normalizing them using main sequence stars.
When the star counts for the
two clusters are normalized near the turnoff, there are more M10 stars
than M12 stars in the magnitude ranges discussed above. 
We can quantify this using the same RGB-MS ratio for the cluster M12
using data from Hargis et al. (2004) since the ``best'' model
parameters are nearly identical between the two clusters. Hargis et
al. found that the M12 LF showed no excess of RGB stars compared to
theory when the LF is normalized to the main sequence. So, as expected
we find $N_{RGB}/N_{MSTO}=0.083\pm0.008$ for the faint sample in M12
($13.95 < I < 14.95$) as compared to 0.0755 from Y$^2$ models (for
[Fe/H] = -1.41, age 12 Gyr, and mass function exponent $x=0$) and
$0.116\pm 0.006$ for M10.  For the brighter sample ($12.15 < I <
13.15$), the ratio is $0.021\pm0.002$ for M12, compared to 0.025 from
Y$^2$ models and 0.0354 for M10.  So, the population ratio values for
M12 agree with theoretical predictions to less than 1 and $2 \sigma$,
while the corresponding ratios for M10 differ by more than 7 and $6
\sigma$, respectively. Thus, M10 has
an excess of RGB stars relative to MS stars when compared to another
well-studied, nearly-identical cluster.

\section{Discussion}\label{disc}

\begin{figure}
\plotone{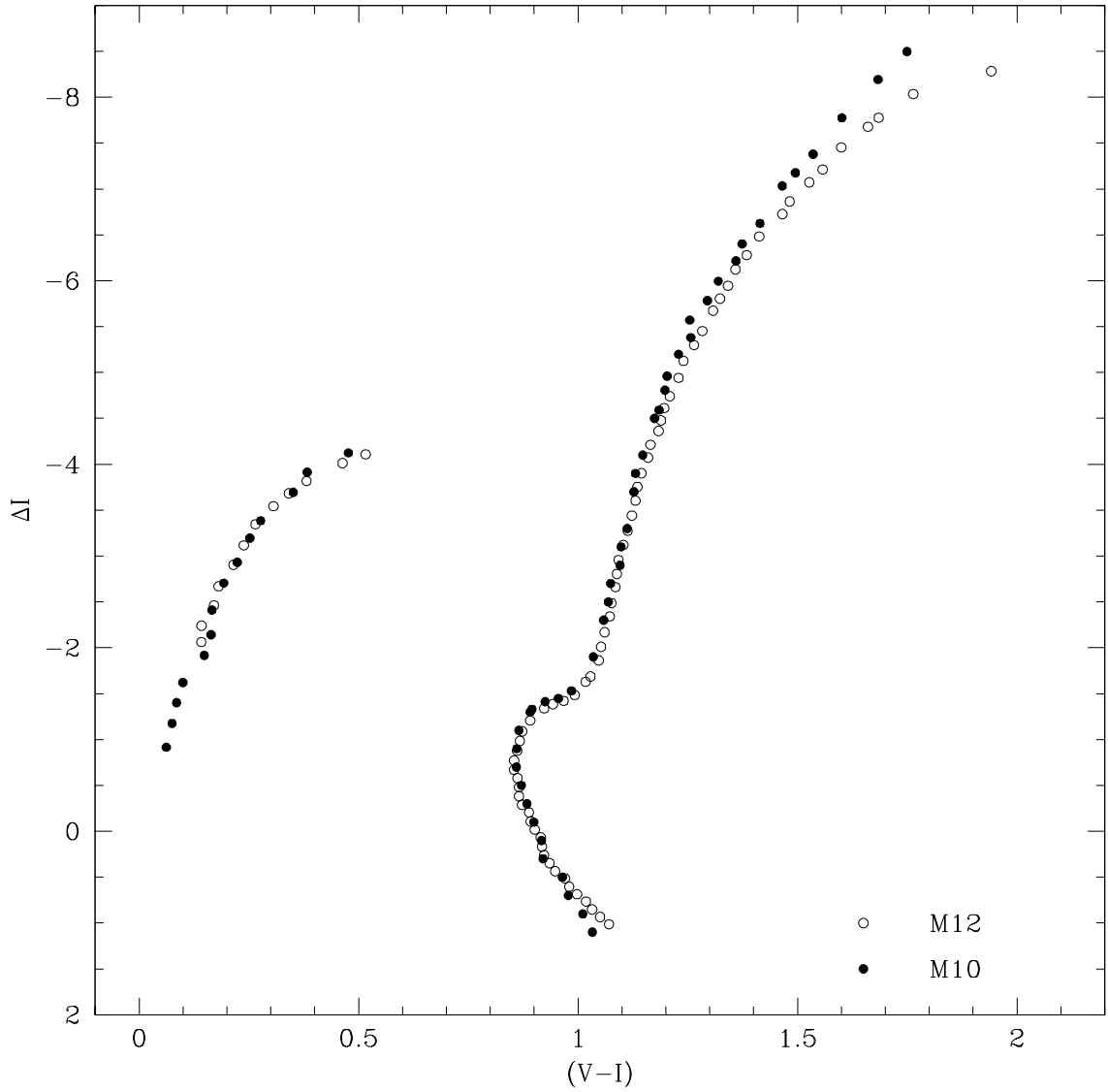}
\caption{Comparison of the fiducial lines of the clusters M10 ({\it open
circles}) and M12 ({\it solid circles}). The turnoff colors of the
fiducials were nearly identical in color, so that the fiducials were
rectified by shifting in magnitude (M10 was 0.107 mag fainter) so that
the points on the main sequence 0.05 mag redder than the turnoff were
aligned.\label{fidcomp}}
\end{figure}

Both the comparison between M10 and theoretical predictions and
between M10 and the nearly identical cluster M12 indicate that the LF
of M10 is anomalous. The RGB-MS
ratio defined above is insensitive to the parameters age and
heavy element content ([Fe/H] and [$\alpha$/Fe] particularly), which
allows us to eliminate them from consideration as the cause.  In
addition to having an age and heavy element content that is identical
to that of M10 to within current observational errors, M12 has a CMD
morphology (most notably a blue horizontal branch tail) that is very
nearly the same as M10. Whatever the cause of M10's unusual LF, it
does not seem to create noticeable differences in the evolutionary
tracks of stars in the CMD except possibly near the RGB tip (see
Fig. \ref{fidcomp}).

\begin{figure}
\plotone{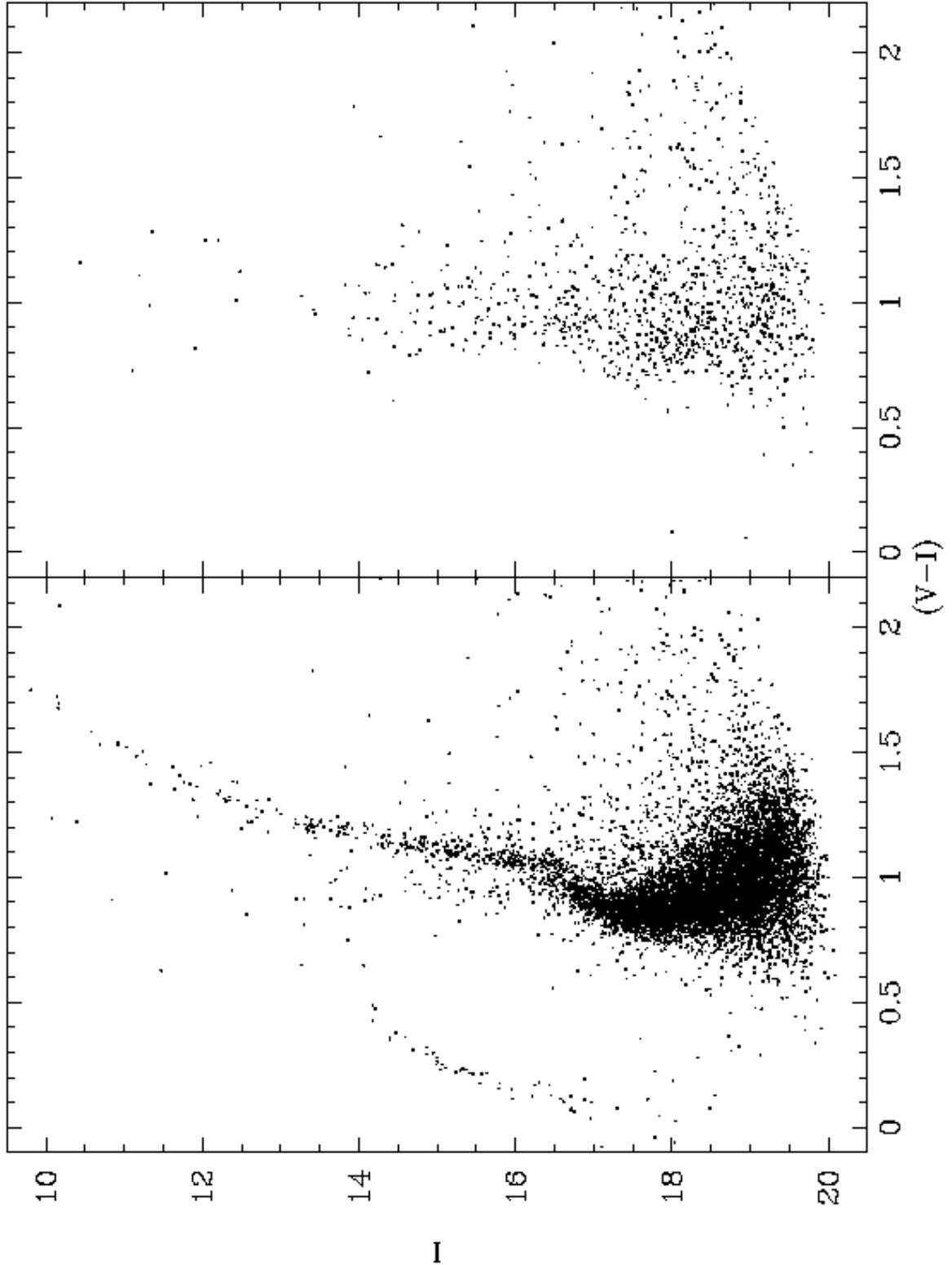}
\caption{{\it Left panel:} Dereddened CMD for M10 stars with
$3\farcm4 < r < 8\farcm5$ from the cluster center.  {\it Right panel:}
Dereddened CMD for stars with $r > 11.\farcm3$ from the cluster
center. Most of these stars are likely to belong to the
field.\label{cmds}}
\end{figure}

\begin{figure}
\plotone{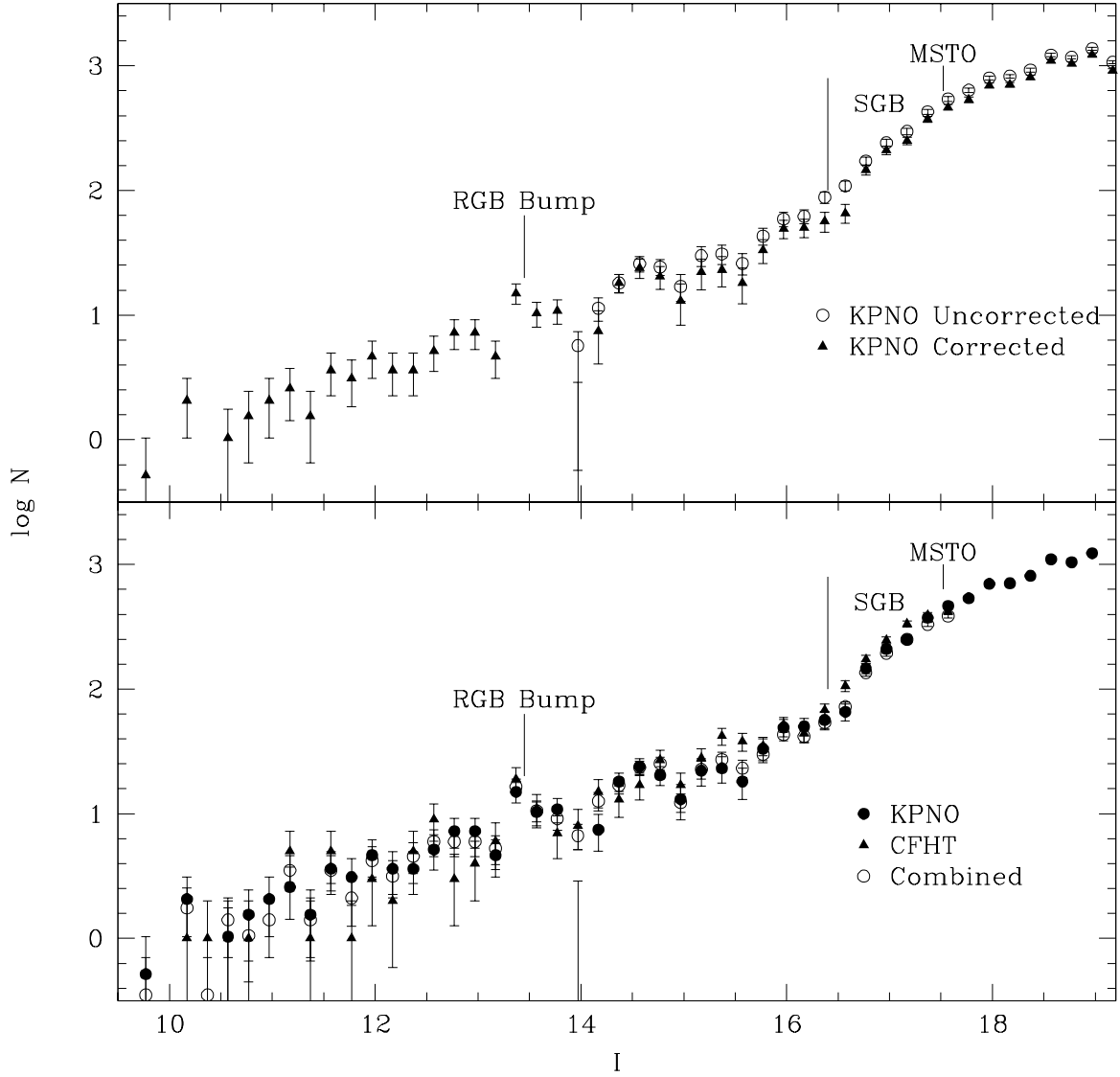}
\caption{{\it Upper panel:} The $I$-band luminosity function with
corrections and without corrections for field star contamination.  No
field stars were detected for $I < 13.9$. {\it Lower panel:}
Comparison of luminosity functions for the KPNO, CFHT, and combined
star samples. \label{fieldcorr}}
\end{figure}

Field-star contamination (or errors involved in correcting for this
contamination) are unlikely to explain the unusual features of the
LF. Field stars are easily visible in the CMD brighter than the
subgiant branch, and can also be detected fainter than the turnoff
based on their colors (significantly redder than the cluster fiducial,
even when the data is dereddened) and based on their lack of any
concentration toward the cluster center (see Fig. \ref{cmds}). Once
corrected, the LF of the main sequence, subgiant branch, and part of
the lower red giant branch agrees very well with theoretical
predictions (see Fig. \ref{fieldcorr}).  However, field star
contamination is not strong enough to explain the LF excess just below
the RGB bump, and contamination is negligible above the bump since the
field star distribution clearly diverges from the RGB for $I \la
15.5$. In addition, because the excess star counts are measured in
both the core {\it and} envelope of the cluster, we can rule out most
other sorts of foreground or background contamination.


It is natural to expect that unusual features in LFs should
result from physics that affects the evolutionary timescales of the
stars.  If that is the case here, we should look at factors that affect
the rate at which nuclear fuel is processed. The abundance of helium
and CNO elements enter into the determination of the instantaneous
nuclear reaction rate at any point in a star, yet theoretical models indicate
that the abundance of CNO elements does {\it not} significantly affect
the relative numbers of RGB and MS stars while helium abundance 
plays a more substantial role \citep{stetlf, larson}. The reason for
this can be seen in the study of \citet{rat} --- the evolutionary
timescale for red giants is {\it not} changed by variations in helium
abundance, but the MS timescale is. We believe that this
can be understood qualitatively as being due to the nature of the
structure of red giants: the rate of nuclear reactions in the hydrogen
fusion shell is set entirely by the star's need to support the
envelope and the position of the shell source (set by the size of the
degenerate core). The structural constraints on giant stars are
responsible for the luminosity -- core mass relations found in
theoretical models of giant stars. When put in this framework, 
the possible causes of the discrepancy seen in M10 are more easily seen.

The evolutionary timescale for red giants can be modified
(independently of the MS) if there is a change in the way the envelope
of the giant is supported. An example of this is seen in the rotating
models of \citet{larson} in which angular momentum was assumed to be
conserved in radiative shells and convection zones were assumed to
rotate like solid bodies. In the red giant stage, the contraction of
the core leads to rapid rotation that provides a modest amount of
centrifugal support to the star's envelope. This relieves the fusion
shell of some of the burden of supporting the envelope (or in other
words, the density of the fusion shell decreases somewhat, causing a
decrease in the energy output of the shell). The degree of centrifugal
support changes as the star evolves thanks to continued contraction of
the core as additional mass is added by the fusion shell. Indirect
evidence for significant core rotation comes from cluster horizontal
branch stars (e.g. Behr 2003): rotational velocities for some of these
stars can reach as high as 40 km s$^{-1}$. Although this is
intriguing, we are far from completely understanding the angular
momentum evolution of giant stars, and there is no reason to believe
that M10 stars have unusually high rotation on average. In addition,
the relatively good agreement of observations and theory on the lower
red giant branch argues against a process that modifies the
evolutionary timescales from the beginning of the red giant branch.

For related reasons, we predict that so-called {\it ``deep mixing'' on
the RGB should not have a significant long-term effect on the
evolutionary timescales of giant stars} unless the extra mixing is
somehow related to an alternate means of envelope support. (``Deep
mixing'' processes are known to produce O-Na anticorrelations at the
surfaces of bright giants in many clusters.) Changes to the chemical
makeup of the material being processed by the fusion shell causes the
shell and envelope of the star to adjust on a timescale much shorter
than the lifetime of the red giant phase. The best-known example of
this is the red giant bump. When the fusion shell reaches what was the
base of the envelope convection zone (in which the gas has a higher
hydrogen content), the star's evolution temporarily slows as it
adjusts to the new fuel mix.  However, the slope of the differential
LF function after the bump is not significantly different from what it
was before, which means that the evolutionary timescales of the RGB
stars are once again enforced by the structure. So, while there could
be modifications to the cluster LF if deep mixing begins at a position
different than the red giant bump, extended periods of deep mixing
would not change the evolutionary timescale once the star's envelope
has adjusted. Spectroscopic analysis of M10 giants \citep{kraft} add
some credence to this idea since M10 giants on average have oxygen
depletions between that of M3 (a cluster that has a
theoretically-predicted ratio of giants to main sequence stars
\citep{rood}) and M13.  So unless the adjustment timescale for the
stellar envelope (approximately the Kelvin-Helmholtz timescale) is
several times longer than that predicted for RGB bump stars evolving
via standard physics {\it and} the star's brightness varies by several
magnitudes during this time, it would be difficult to explain M10's LF
in this way.

Ratios of giants to main sequence stars are obviously a function of
the evolutionary timescales of both main sequence stars and
giants. For MS stars, differences in initial helium abundance cause
related differences in evolutionary timescale for the simple reason
that the amount of fuel changes (although there are additional effects
resulting from changes to the star's luminosity). We do not believe
that differences in initial helium content are to blame because i) a
large positive enhancement would be necessary for M10, bringing M10's
helium content close to the solar value and going against expected
nucleosynthetic trends ($\Delta Y / \Delta Z > 0$) \citep{larson}, ii)
evidence from the helium abundance indicator $R$ shows that the
globular cluster system as a whole does not have an intrinsic spread
in $Y$ of greater than 0.019 \citep{salaris}, iii) there is virtually
no difference in the CMD morphologies of M10 and M12, and iv) a change
in main sequence timescales alone would tend to make the RGB star
counts higher than predicted at all luminosity levels.

The overall insensitivity of the ratio $N_{RGB} / N_{MS}$ to most
parameters that vary from cluster to cluster makes M10's LF difficult
to understand. The theoretical results do help to emphasize that in
spite of the substantial structural differences between MS and RGB
stars, there is a very strong connection between their evolutionary
timescales.  However, we must also keep in mind that clusters are made
of individual stars. From the time a star reaches the turnoff of a
globular cluster, it can take approximately 2 Gyr to reach the middle
of the RGB \citep{yitracks}. While the mass function of {\it current}
main sequence stars can be estimated from the slope of the LF, the
mass function of now-evolved stars cannot --- in fact, it may be
different.  Typically, theoretical LFs assume that the slope of
the initial mass function for stars in a cluster is a constant.
If this is not the case, then $N_{RGB} / N_{MS}$ could differ from
its well-determined theoretical value.

The question remains whether such a variation in the initial mass
function could be due to a statistical fluctuation or whether it would
have to result from a stronger trend in the mass function of cluster
stars. For $12.2 < I < 13.2$, 79 stars were identified while only
about 51 were expected. For $14 < I < 15$, 250 stars were identified,
and 149 were expected theoretically. Poisson fluctuations ($\sqrt{N}$)
in ``normal'' samples of stars as a function of mass therefore seem
unlikely to account for the M10 observations. All of the stars
currently on the giant branch would have originated from just one LF
bin at the turnoff approximately 2 Gyr ago, but even Poisson
fluctuations in a sample that size ($< 1000$ stars) are several times
too small to explain the giant branch excesses we see. 

Another possibility is that the mass function at the cluster turnoff
about 2 Gyr ago deviated more systematically from the power-law form
assumed in models. Although this hypothesis cannot be tested directly, we can
look for signs of significant variations in mass function slope on the
current main sequence. We compared the bin-to-bin LF slopes with
theoretical values using the best ``average'' mass function exponent
$x = 2.0$. On the upper main sequence, several bins near the turnoff
deviate from the theoretical LF slopes by between 2 and $3\sigma$, and
one interval ($18.57 < I < 18.77$) closer to our faint limit deviates
by almost $4.3 \sigma$.

Of the main sequence LFs for M10, M22, and M55 discussed by
\citet{Piotto99}, M10's LF was least consistent with having a constant
mass function slope $x$ (although there is significant uncertainty in the
mass-luminosity relation for the low mass stars). We note once again
that an examination of the LF \citep{Piotto99} indicates that there is
a noticeable drop in the main sequence luminosity function for $21 \la
V \la 23$ (it is less obvious in $I$), so that it is not
unreasonable to believe that there might have been another mass
function variation among stars that have now left the main sequence.
The anomaly seen in our LF of the evolved stars of M10 implies that
there may have been a greater number of MS stars than expected from
the $x = 2$ mass function derived from the upper main sequence LF.


A simple understanding of dynamical processes (like evaporation) would
lead us to expect that strong dynamical effects would tend to remove
low mass stars from a cluster, making the mass function
shallower. Because the main sequence LF observed by \citet{Piotto99}
is much steeper than those of M22 and M55 (but comparable to the
metal-poorer clusters M15, M30, and M92), the implication is that M10
has been affected in a comparatively minor way. Even if dynamics are
responsible for modifying M10's LF, we are again left to explain why
M12 does not show a similar effect, given that the characteristics of
its Galactic orbit are similar to that of M10 \citep{dine}.

In the end, we do not have a satisfying explanation for the unusual
aspects of M10's LF, but the cause can be external to the stars
themselves. Though there is evidence for mass function variations
(changes in mass function slope) from other parts of M10's LF, we
still have no explanation for why they might be present in a massive
cluster like M10. Although the hypothesis is difficult to test, there
are two possible avenues to follow. First, continued searches for
unusual RGB luminosity functions in other globular clusters may turn
up additional examples that would help to identify a relationship to
cluster parameters (although the most obvious correlations with
metallicity or horizontal branch morphology seem to be ruled out by
previously published LFs) or identify whether RGB anomalies cover
similar or different ranges of luminosity. Second, main sequence
luminosity functions using larger samples would help answer the
question of whether significant (non-Poissonian) LF fluctuations exist
within clusters. Most deep MS LFs have been derived using the WFPC2
imager on the {\it Hubble Space Telescope}. Wide-field high
spatial-resolution instruments (like the Advanced Camera for Surveys
on HST or cameras on the CFHT) would be ideal for surveying the larger
areas necessary.

\acknowledgments We would like to thank J. Hesser for his role in
getting the CFHT observations, and the anonymous referee for comments
that have strengthened the manuscript. E.L.S. would like to thank
J. Faulkner for (long-ago) conversations about red giants, and R. Rood
for pointing out the potential role of LF fluctuations. This work has
been funded through grant AST 00-98696 from the National Science
Foundation to E.L.S. and M.B.

\begin{deluxetable}{lrr}
\tablewidth{0pt}
\tablecaption{M10 $(I,V-I)$ Fiducial Points}
\tablehead{\colhead{$I$} & \colhead{$V-I$} & \colhead{$N$}}
\startdata
19.4000 &  1.0327 &  1072 \\
19.2000 &  1.0109 &  1353 \\
19.0000 &  0.9778 &  1531 \\
18.8000 &  0.9642 &  1536 \\
18.6000 &  0.9203 &  1470 \\
18.4000 &  0.9168 &  1427 \\
18.2000 &  0.8994 &  1391 \\
18.0000 &  0.8837 &  1329 \\
17.8000 &  0.8710 &  1140 \\
17.6000 &  0.8594 &  1067 \\
\enddata
\label{fidtab}
\tablecomments{The complete version of this table is in the
electronic edition of the Journal. The printed edition contains only a sample.}
\end{deluxetable}

\begin{deluxetable}{rrrrrrrr}
\tablewidth{0pt}
\tablecaption{M10 $I$-band Luminosity Function}
\tablehead{\colhead{$I$} & \colhead{$N$} & \colhead{$\sigma_N$} & \colhead{$N$}
& \colhead{$\sigma_N$} & \colhead{$N$} & \colhead{$N$} & \colhead{$\sigma_N$} \\
 & \multicolumn{2}{c}{KPNO Uncorrected} & \multicolumn{2}{c}{Field Star
 Corrected} & \colhead{CFHT} & \multicolumn{2}{c}{Combined}\\}
\startdata
 9.77 & 0.517 & 0.517 & 0.517 & 0.517 & 0 & 0.342 & 0.342 \\
10.17 & 2.067 & 1.033 & 2.067 & 1.033 & 1 & 1.709 & 0.764 \\
10.37 & 0.000 &       & 0.000 &       & 1 & 0.342 & 0.342 \\
10.57 & 1.033 & 0.731 & 1.033 & 0.731 & 2 & 1.367 & 0.683 \\
10.77 & 1.550 & 0.895 & 1.550 & 0.895 & 1 & 1.025 & 0.592 \\
10.97 & 2.067 & 1.033 & 2.067 & 1.033 & 0 & 1.367 & 0.683 \\
11.17 & 2.583 & 1.155 & 2.583 & 1.155 & 5 & 3.417 & 1.081 \\
11.37 & 1.550 & 0.895 & 1.550 & 0.895 & 1 & 1.367 & 0.683 \\
11.57 & 3.617 & 1.367 & 3.617 & 1.367 & 5 & 3.417 & 1.081 \\
11.77 & 3.100 & 1.266 & 3.100 & 1.266 & 1 & 2.050 & 0.837 \\
11.97 & 4.651 & 1.550 & 4.651 & 1.550 & 3 & 4.101 & 1.184 \\
\enddata
\label{lftab}
\tablecomments{The complete version of this table is in the
electronic edition of the Journal. The printed edition contains only a sample.}
\end{deluxetable}
\end{document}